\documentclass[amsmath,amssymb,showpacs,twocolumn]{revtex4-1}
\usepackage{graphicx,bm,color,subfigure}

\begin{document}

\title{Doping-type-dependent pairing symmetry in predicted Ni-based high-$T_c$ superconductor La$_2$Ni$_2$Se$_2$O$_3$}

\author{Li-Da Zhang}
\email{zhangld@bit.edu.cn}
\affiliation{School of Physics, Beijing Institute of Technology, Beijing 100081, China}

\begin{abstract}
We study the electronic instabilities of newly predicted Ni-based high-$T_c$ superconducting material La$_2$Ni$_2$Se$_2$O$_3$ based on the random phase approximation. Our calculations on the susceptibility indicate that the collinear antiferromagnetic state in the parental compound is induced by the perfect Fermi surface nesting. Our further calculations reveal that the ground states of the doped compound are the $s_{\pm}$- and $d_{xy}$-wave superconducting states driven by the antiferromagnetic spin fluctuations enhanced by the quasi-nestings. Interestingly, the $s$- and $d$-wave pairings occur in the hole and electron doping cases, respectively. This doping-type dependence of the pairing symmetry can be understood from the doping dependence of the nested Fermi pockets.
\end{abstract}



\maketitle


\section{Introduction}
Cuprates and iron-based superconductors, as only two families of discovered high-$T_c$ superconducting (SC) materials, share common properties in many aspects. Recently, the efforts to unify these two families theoretically reveal a hidden quasi-two-dimensional electronic structure shared by the two families \cite{Hu1,Hu2}. In this electronic environment, the $d$-orbitals of transition metal atoms with the strongest in-plane coupling to the $p$-orbitals of anions are isolated near Fermi energy. Using this condition as a guide to search for or design new high-$T_c$ SC materials has given rise to a candidate configuration with $d^7$ filling, hosted mainly by the Co$^{2+}$-based compounds \cite{Hu1,Hu3,Hu4}.

In Ref. \cite{Hu5}, a new candidate configuration with $d^8$ filling hosted by the Ni$^{2+}$-based compounds is proposed. Particularly, as the prototype of this Ni$^{2+}$-based family, the layered compounds La$_2$Ni$_2$M$_2$O$_3$ (M=S, Se, Te) are analyzed in details. It is found that the antiferromagnetic (AFM) superexchange interaction is maximized to form a collinear AFM state in the parental compound. For the doped compounds, there is a strong competition between the $s$-wave and $d$-wave SC states. Specifically, while the extended $s$-wave is favored upon hole doping, the $d$-wave can become highly competitive under electron doping. The above results are obtained by the analysis based on the energy scale and general principle emerged in understanding both cuprates and iron-based superconductors. Thus, the quantitative results, especially the ground states and phase diagrams of the system, still need to be explored.

In this paper, we investigate the magnetism and the superconductivity in La$_2$Ni$_2$Se$_2$O$_3$ for different doping levels and interaction parameters to determine the ground states and phase diagrams of the system. We adopt the four-orbital tight-band (TB) model proposed in Ref. \cite{Hu5} to describe the band structure of single layer Ni$_2$Se$_2$O in La$_2$Ni$_2$Se$_2$O$_3$, and the Hubbard-Hund model to further describe the interactions in the system. Our calculation based on the random phase approximation (RPA) confirms that, for the realistic interaction parameters, the undoped compound hosts an collinear AFM state as predicted in Ref. \cite{Hu5}. In particular, we find that the AFM state is induced by the perfect Fermi surface (FS) nesting. As for doped compounds, we identify the $s_{\pm}$- and $d_{xy}$-wave SC states as the ground states of the system upon hole and electron dopings, respectively. Although both the $s$- and $d$-wave pairings are mediated by the AFM spin fluctuations enhanced by the quasi-nestings, the specific nestings for these two pairings come from different Fermi pockets. Specifically, while the pockets for the $s$-wave pairing keep large enough to support the nesting under all doping levels, most pockets for the $d$-wave pairing grow large enough to support the nesting only under relative high electron doping levels. This accounts for the emergence of the $d$-wave pairing under electron doping naturally. We stress that the results we present here are robust against the changes of the interaction parameters as well as the doping level (with fixed doping type). Thus, we can expect that the ground states and phase diagrams of the system are suitable for the other Ni-based high-$T_c$ superconductors which share the similar single layer Ni$_2$M$_2$O (M=S, Se, Te).

The rest of this paper is organized as follows. In Sec. II, we describe the effective four-orbital TB model of single layer Ni$_2$Se$_2$O constructed in Ref. \cite{Hu5}, and analyze the bare susceptibility of the system. In Sec. III, we study the SC pairing symmetry as well as the magnetic order of the system, and depict the ground states and phase diagrams for different doping levels and interaction parameters. Finally, in Sec. IV, a conclusion will be reached.

\section{TB Model}
\begin{figure}[htbp]
\centering
\includegraphics[width=0.48\textwidth]{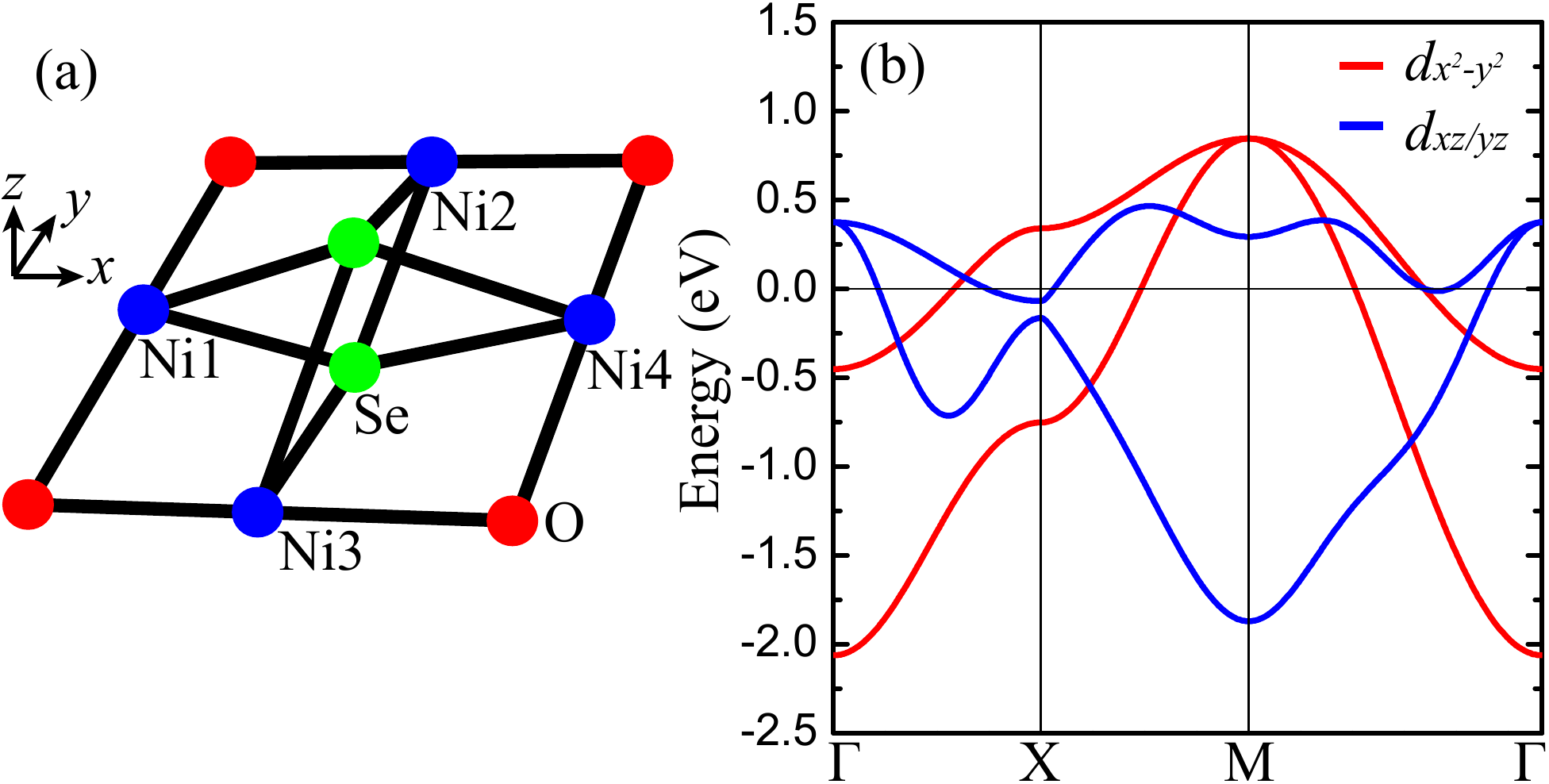}
\caption{(a) The lattice structure of the single [Ni$_2$Se$_2$O]$^{2-}$ layer in a unit cell. (b) The band structure of the TB model with orbital characters.}\label{bands}
\end{figure}

The active ingredient of the layered superconductor La$_2$Ni$_2$Se$_2$O$_3$ is the [Ni$_2$Se$_2$O]$^{2-}$ layer, which has the lattice structure shown in Fig. \ref{bands}(a). The two $e_g$ orbitals of Ni cations determine the low energy physics of the system. In the basis of the four $e_g$ orbitals at two different Ni sites (Ni3 $d_{x^2-y^2}$, Ni3 $d_{yz}$, Ni4 $d_{x^2-y^2}$, Ni4 $d_{xz}$), a minimum effective TB Hamiltonian, $H_0$, are constructed in Ref. \cite{Hu5} to capture the band structure near FS of the single layer Ni$_2$Se$_2$O. The nonzero elements of the matrix of the TB Hamiltonian are written as
\begin{align}\label{hop}
&H_{11}(k_x,k_y)=\epsilon_1+2t^{11}_{xx}\cos(k_x)+2t^{11}_{yy}\cos(k_y),    \nonumber\\
&H_{22}(k_x,k_y)=\epsilon_2+2t^{22}_{yy}\cos(k_y)+2t^{22}_{yyyy}\cos(2k_y)  \nonumber\\
&~~~~+4t^{22}_{xxyy}\cos(k_x)\cos(k_y)+4t^{22}_{yyyyxx}\cos(k_x)\cos(2k_y), \nonumber\\
&H_{33}(k_x,k_y)=H_{11}(k_y,k_x),                                           \nonumber\\
&H_{44}(k_x,k_y)=H_{22}(k_y,k_x),                                           \nonumber\\
&H_{13}=H_{31}=4t^{13}_{xy}\cos(0.5k_x)\cos(0.5k_y),                        \nonumber\\
&H_{24}=H_{42}=-4t^{24}_{xy}\sin(0.5k_x)\sin(0.5k_y).
\end{align}
Here, in unit of eV, the onsite energy $\epsilon_1=7.2218$, $\epsilon_2=7.0804$, and the hopping parameters $t^{11}_{xx}=-0.3995$, $t^{11}_{yy}=-0.1264$, $t^{22}_{yy}=0.1573$, $t^{22}_{yyyy}=0.0656$, $t^{22}_{xxyy}=-0.0113$, $t^{22}_{yyyyxx}=0.0668$, $t^{13}_{xy}=-0.2014$, $t^{24}_{xy}=-0.2705$. In Fig. \ref{bands}(b), we show the band structure of the TB model with half-filling chemical potential $7.3744$ eV, which corresponds to the undoped case.

\begin{figure}[htbp]
\centering
\includegraphics[width=0.45\textwidth]{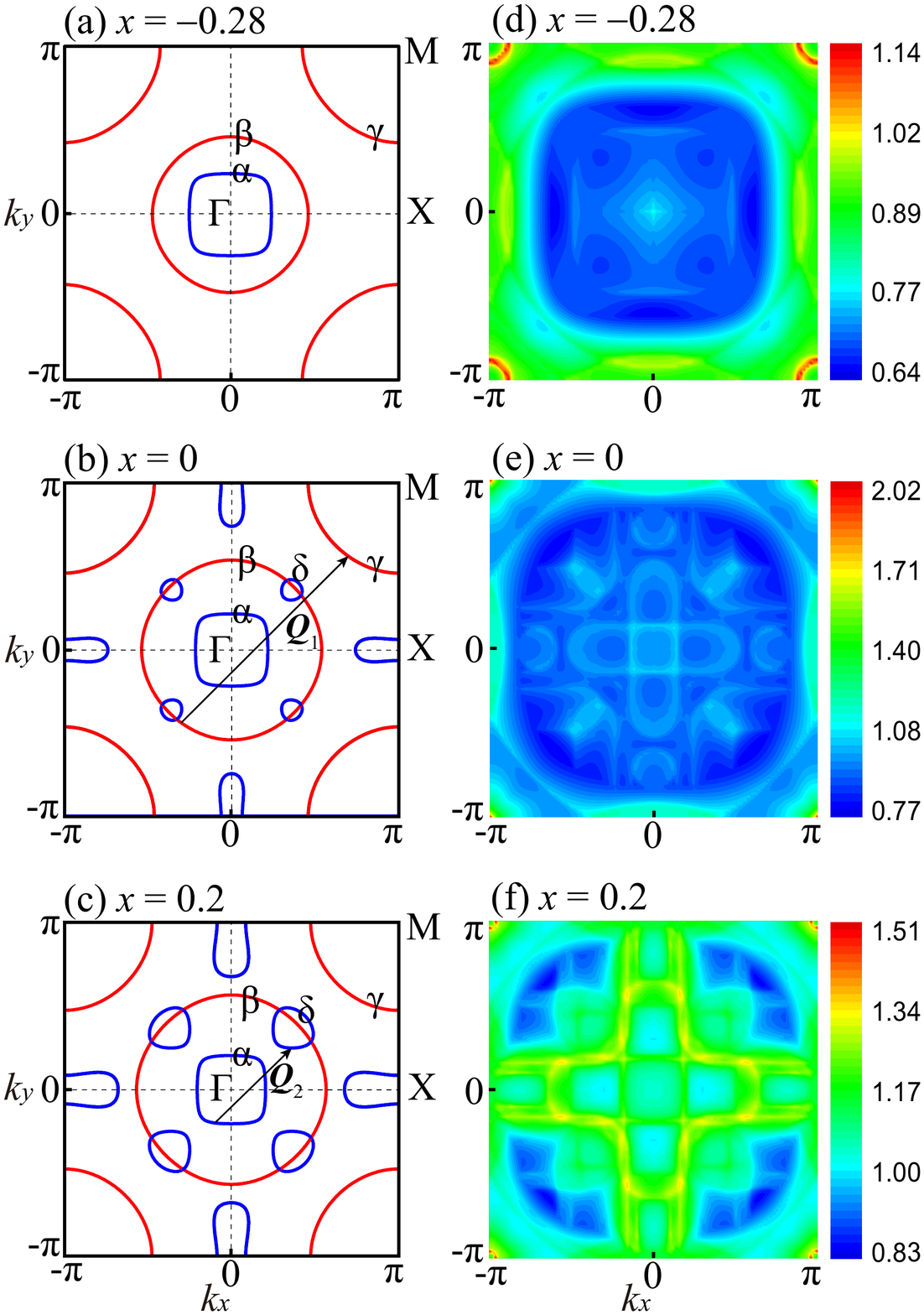}
\caption{(a)-(c) The FSs under different doping levels $x$ studied in Ref. \cite{Hu5}. The red (blue) Fermi pockets are from the bands with $d_{x^2-y^2}(d_{xz/yz})$-orbital character. $\bm{Q}_1$ and $\bm{Q}_2$ are the nesting vectors connecting the Fermi pockets. The equivalent nesting vectors under the $90^o$ rotation are not shown here. (d)-(f) The $\bm{k}$-space distributions of the largest eigenvalue of susceptibility matrix $\chi^{(0)pp}_{qq}(\bm{k},i\omega_n=0)$ for the doping levels corresponding to (a)-(c).}\label{FSchi}
\end{figure}

We define the doping level $x\equiv n_e-4$ with $n_e$ being the number of electrons in the basis orbitals, so that the negative and positive $x$ represent the hole and electron dopings, respectively. In Figs. \ref{FSchi}(a)-\ref{FSchi}(c), we depict the FSs of the system at doping levels studied in Ref. \cite{Hu5}. As the band structure is decoupled between $d_{x^2-y^2}$ and $d_{xz/yz}$ orbitals as shown in Fig. \ref{bands}(b), the FSs are decoupled correspondingly. For the undoped case shown in Fig. \ref{FSchi}(b), two half-filling $d_{x^2-y^2}$ orbitals give rise to the perfectly nested electron pocket $\beta$ and hole pocket $\gamma$ with nesting vector $\bm{Q}_1=(\pi,\pi)$. For the doped cases shown in Figs. \ref{FSchi}(a) and \ref{FSchi}(c), the nearly perfect nestings between pockets $\beta$ and $\gamma$ exist as well, which implies the weak doping dependence of these two pockets. Also, for the electron-doped case shown in Fig. \ref{FSchi}(c), there exists the nestings between the pockets with $d_{xz/yz}$-orbital character, such as the one between pockets $\alpha$ and $\delta$ with nesting vector $\bm{Q}_2$.

To investigate the ordered states of the system, we adopt the multi-orbital RPA approach \cite{RPA1,RPA2,RPA3,RPA4,Scalapino1,Scalapino2,Liu2013,Wu2014,Ma2014,Zhang2015}. Specifically, we first define the bare susceptibility
\begin{align}\label{chi0}
\chi^{(0)pq}_{st}(\bm{k},\tau)\equiv
&\frac{1}{N}\sum_{\bm{k}_1\bm{k}_2}\left\langle
T_{\tau}c_{p}^{\dagger}(\bm{k}_1,\tau)
c_{q}(\bm{k}_1+\bm{k},\tau)\right.                      \nonumber\\
&\left.\times c_{s}^{\dagger}(\bm{k}_2+\bm{k},0)
c_{t}(\bm{k}_2,0)\right\rangle_0.
\end{align}
Here $\langle\cdots\rangle_0$ denotes the thermal average for the noninteracting system, $T_{\tau}$ denotes the time-ordered product, and $p,q,s,t=1,2,3,4$ are the orbital indices. Fourier transformed to the imaginary frequency space, the bare susceptibility can be expressed as
\begin{align}\label{chi0e}
\chi^{(0)pq}_{st}(\bm{k},i\omega_n)
=&\frac{1}{N}\sum_{\bm{k}'ab}
\xi^{a}_{t}(\bm{k}')
\xi^{a*}_{p}(\bm{k}')
\xi^{b}_{q}(\bm{k}'+\bm{k})                         \nonumber\\
&\times\xi^{b*}_{s}(\bm{k}'+\bm{k})
\frac{n_F(\varepsilon^{b}_{\bm{k}'+\bm{k}})
-n_F(\varepsilon^{a}_{\bm{k}'})}
{i\omega_n+\varepsilon^{a}_{\bm{k}'}
-\varepsilon^{b}_{\bm{k}'+\bm{k}}}.
\end{align}
Here $a,b=1,2,3,4$ are band indices, $\varepsilon^{a}_{\bm{k}}$ and $\xi^{a}\left(\bm{k}\right)$ are the $a$-th eigenvalue and eigenvector
of the TB Hamiltonian, respectively, and $n_F$ is the Fermi-Dirac distribution
function.

In Figs. \ref{FSchi}(d)-\ref{FSchi}(f), we depict the $\bm{k}$-space distributions of the largest eigenvalue of susceptibility matrix $\chi^{(0)pp}_{ss}(\bm{k},i\omega_n=0)$ for different doping levels. For the undoped case shown in Fig. \ref{FSchi}(e), the perfect nesting with wavevector $\bm{Q}_1=(\pi,\pi)$ leads to the largest eigenvalue peaking perfectly at the M point, which is in consistent with the AFM state of the parental compound predicted in Ref. \cite{Hu5}. For the doped cases shown in Figs. \ref{FSchi}(d) and \ref{FSchi}(f), the nearly perfect $(\pi,\pi)$ nestings lead to largest eigenvalues peaking around the M point, which implies that the $(\pi,\pi)$ AFM spin correlations are important for both hole and electron dopings. Also, for the electron doping case shown in Fig. \ref{FSchi}(f), the nestings with wavevectors connecting $d_{xz/yz}$-orbital pockets, such as $\bm{Q}_2$, lead to the subpeaks far away from the M point. These subpeaks have the intensities comparable to those of the peaks around the M point, and thus would bring new physics other than those induced by the $(\pi,\pi)$ nesting.

\section{AFM and SC states}
To reveal the various instabilities of the system, we further consider the interactions described by the following Hubbard-Hund Hamiltonian:
\begin{align}\label{model}
H=&H_0+H_{int}\nonumber\\
H_{int}=&U\sum_{i\mu}n_{ip\uparrow}n_{ip\downarrow}+
V\sum_{i,p<q}n_{ip}n_{iq}+J\sum_{i,p<q}                   \nonumber\\
&\Big[\sum_{\sigma\sigma^{\prime}}c^{+}_{ip\sigma}c^{+}_{iq\sigma^{\prime}}
c_{ip\sigma^{\prime}}c_{iq\sigma}+(c^{+}_{ip\uparrow}c^{+}_{ip\downarrow}
c_{iq\downarrow}c_{iq\uparrow}+h.c.)\Big].
\end{align}
Here, $i$ is the site index, and $p,q=1,2$ are the orbital indices on a site. General symmetry argument requires $U=V+2J$, and thus, we adopt $U$ and $J/U$ as independent parameters.

When the above interactions is turned on, we can further define the spin $(s)$ and charge $(c)$ susceptibilities. In the RPA level, these renormalized susceptibilities are expressed in the form of Dyson-type matrix equations as
\begin{align}\label{chisce}
\chi^{(s,c)}(\bm{k},i\omega_n)=[I\mp\chi^{(0)}(\bm{k},i\omega_n)
U^{(s,c)}]^{-1}\chi^{(0)}(\bm{k},i\omega_n).
\end{align}
Here $\chi^{(s,c)}(\bm{k},i\omega_n)$, $\chi^{(0)}(\bm{k},i\omega_n)$, and $U^{(s,c)}$ are operated as $4^2\times 4^2$ matrices. The nonzero elements of the matrix $U^{(s,c)}$ are
\begin{align}\label{usc}
&U^{(s)pq}_{st}=\left\{
\begin{array}{cc}
U,   & p=q=s=t;     \\
J_H, & p=q\neq s=t; \\
J_H, & p=s\neq q=t; \\
V,   & p=t\neq q=s,
\end{array}
\right.                   \\
&U^{(c)pq}_{st}=\left\{
\begin{array}{cc}
U,      & p=q=s=t;     \\
2V-J_H, & p=q\neq s=t; \\
J_H,    & p=s\neq q=t; \\
2J_H-V, & p=t\neq q=s,
\end{array}
\right.                   \\
&U^{(s,c)p+2,q+2}_{s+2,t+2}=U^{(s,c)pq}_{st},
\end{align}
where $p,q,s,t=1,2$.

From Eq. (\ref{chisce}), the spin (charge) susceptibility will be enhanced (suppressed) as the repulsive Hubbard interaction increases. In particular, the spin susceptibility will diverge at the very $\bm{k}$-points where the bare susceptibility matrix $\chi^{(0)pp}_{ss}(\bm{k},i\omega_n=0)$ peaks, when $U$ reaches a critical interaction strength $U_c$ for a fixed $J/U$ and $i\omega_n=0$. This divergence implies the formation of the long-range magnetic order. For the undoped case, as the divergence occurs perfectly at the M point, the magnetic order is collinear $(\pi,\pi)$ AFM as predicted in Ref. \cite{Hu5}.

\begin{figure}[htbp]
\centering
\includegraphics[width=0.4\textwidth]{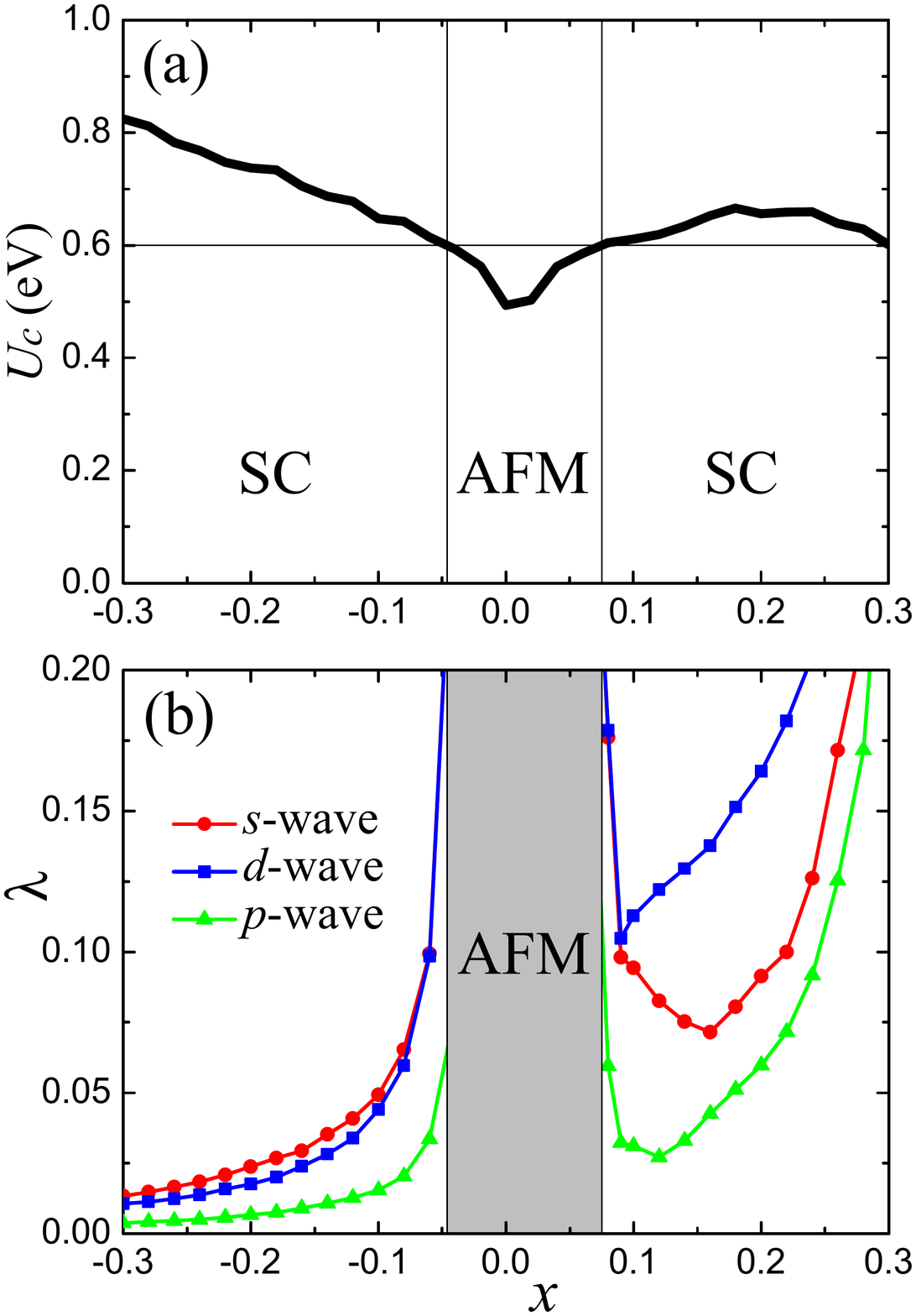}
\caption{(a) The doping dependence of the critical interaction $U_c$. The horizontal line indicates $U=0.6$ eV adopted in our calculations, and the vertical lines indicate the corresponding phase boundary between AFM and SC phases. (b) The doping dependence of the largest eigenvalues $\lambda$ of $s$-, $d$-, and $p$-wave SC states for $U=0.6$ eV and $J/U=0.1$.}\label{UcSC}
\end{figure}

In Fig. \ref{UcSC}(a), we further depict the doping dependence of $U_c$ for realistic $J/U=0.1$. Clearly, the $U_c$ has low values around the half filling, and relative high values at heavy doping levels no mater the hole or electron doping. This doping dependence of $U_c$ can be understood from Eq. (\ref{chisce}) and the degree of perfection of the FS nesting. Specifically, a more perfect nesting will enhance the intensity of the bare susceptibility $\chi^{(0)}$, and subsequently, suppress $U_c$ according to Eq. (\ref{chisce}). For example, the perfect nesting between the Fermi pockets $\beta$ and $\gamma$ [see Fig. \ref{FSchi}(b)] leads to suppressed $U_c$ around the half filling, and the approaching perfect nestings between the $d_{xy}$-orbital pockets lead to suppressed $U_c$ near the high doping level $x=0.3$.

The suppression of $U_c$ around the half filling implies that the long-range AFM state occurs more readily in the undoped compound than in the doped one. For a fixed $U$, like $U=0.6$ eV, we can determine the low doping range where $U>U_c$, and thus, the AFM state serves as the leading instability, as shown in Fig. \ref{UcSC}(a). Out of this doping range, we have $U<U_c$, and the possible SC states mediated by the AFM spin fluctuations as the leading instabilities. This doping dependence of the instability is similar to those of cuprates and iron-based superconductors.

To investigate the SC instability occurring when $U<U_c$, we consider a Cooper pair scatter from the state $(\bm{k}',-\bm{k}')$ in the $b$-th band to the state $(\bm{k},-\bm{k})$ in the $a$-th band via exchanging spin fluctuations. This scatter can be describe by the following effective interaction vertex \cite{Scalapino1,Scalapino2}:
\begin{align}\label{vv}
V^{ab}(\bm{k},\bm{k}')=
{\rm Re}\sum_{pqst}\Gamma^{pq}_{st}(\bm{k},\bm{k}')
\xi^{a*}_{p}(\bm{k})                            \nonumber\\
\xi^{a*}_{q}(-\bm{k})
\xi^{b}_{s}(-\bm{k}')
\xi^{b}_{t}(\bm{k}'),
\end{align}
where $p,q,s,t=1,2,3,4$. For the singlet pairing channel, $\Gamma^{pq}_{st}(\bm{k},\bm{k}')$ take the following symmetrized form:
\begin{align}\label{gams}
\Gamma&^{pq}_{st}(\bm{k},\bm{k}')
=\frac{1}{4}(U^{(c)}+3U^{(s)})^{pt}_{qs}                            \nonumber\\
&+\frac{1}{4}[3U^{(s)}\chi^{(s)}U^{(s)}
-U^{(c)}\chi^{(c)}U^{(c)}]^{pt}_{qs}(\bm{k}-\bm{k}')                \nonumber\\
&+\frac{1}{4}[3U^{(s)}\chi^{(s)}U^{(s)}
-U^{(c)}\chi^{(c)}U^{(c)}]^{ps}_{qt}(\bm{k}+\bm{k}'),
\end{align}
and for the triplet pairing channel, the following antisymmetrized form:
\begin{align}\label{gamt}
\Gamma&^{pq}_{st}(\bm{k},\bm{k}')
=\frac{1}{4}(U^{(c)}-U^{(s)})^{pt}_{qs}                             \nonumber\\
&-\frac{1}{4}[U^{(s)}\chi^{(s)}U^{(s)}
+U^{(c)}\chi^{(c)}U^{(c)}]^{pt}_{qs}(\bm{k}-\bm{k}')                \nonumber\\
&+\frac{1}{4}[U^{(s)}\chi^{(s)}U^{(s)}
+U^{(c)}\chi^{(c)}U^{(c)}]^{ps}_{qt}(\bm{k}+\bm{k}').
\end{align}
Note that, in Eqs. (\ref{vv}), (\ref{gams}) and (\ref{gamt}), we have taken static limit $\omega_n=0$.

From the effective interaction vertex (\ref{vv}), we can obtain the following linearized gap equation:
\begin{align}\label{gapeq}
-\frac{1}{(2\pi)^2}\sum_{b}\oint_{FS}
d^{2}\bm{k}'_{\Vert}\frac{V^{ab}(\bm{k},\bm{k}')}
{v^{b}_{F}(\bm{k}')}\Delta_{b}(\bm{k}')=\lambda
\Delta_{a}(\bm{k}).
\end{align}
Here the integration is along the Fermi pockets labeled by the band index $b$, $v^{b}_F(\bm{k}')$ is the Fermi velocity, and $\bm{k}'_\parallel$ is the component of $\bm{k}'$ along the FS. Solving this gap equation as an eigenvalue problem, one can obtain the pairing eigenvalue $\lambda$ and the corresponding eigenvector $\Delta_a(\bm{k})$ as the relative gap function in different pairing channels. The SC critical temperature $T_c$ is related to the largest $\lambda$ through $T_c\propto\exp(-1/\lambda)$. The leading pairing symmetry is determined by the $\Delta_a(\bm{k})$ corresponding to the largest $\lambda$.

The tetragonal point group of the system determines that, in the singlet channel, the pairing symmetries mainly include $s$- and $d$-wave ones, which have the gap functions remaining and changing the sign under every $90^o$ rotation, respectively. On the other hand, in the triplet channel, the pairing symmetry is mainly the $p$-wave one with doubly degenerate gap functions, which usually mix into the $p+ip$ pairing under $T_c$ to minimize the energy. For the present system, as the main spin fluctuations are AFM, which favors the singlet pairing rather than the triplet, the $p$-wave pairing is hardly to be the leading one.

To determine the specific pairing symmetry at different doping levels, we solve the gap equation (\ref{gapeq}), and depict doping dependence of the largest eigenvalues $\lambda$ for different pairing symmetries with fixed $U=0.6$ eV and $J/U=0.1$ in Fig. \ref{UcSC}(b). A few features revealed here are as follows. First, the most interesting feature is that the $s$- and $d$-wave pairing states dominate in the hole and electron doping cases, respectively. The explanation for this feature involves the distributions of the SC gap functions on the Fermi pockets, which we will present later on. Second, it is obvious that the electron doping side has higher $\lambda$ than the hole doping side, and thus has higher $T_c$. This can be easily understood from the fact that the electron doping side has lower $U_c$, and thus, the same $U$ would lead to the stronger spin fluctuations in the electron doping case. Third, we can see that the eigenvalues $\lambda$ of all SC states tend to diverge in the doping ranges near the phase boundaries between AFM and SC phases. Although this divergence is an artifact in the RPA caused by ignorance of the renormalization of the single particle Green¡¯s function, the stronger AFM spin fluctuations in these critical doping ranges would indeed favor the formation of the superconductivity with higher $T_c$. Finally, the eigenvalues $\lambda$ of the $s$- and $d$-wave SC states are closer to each other when the doping level approaches the AFM and SC phase boundaries from the SC sides. This feature favors the formation of the $s+id$ pairing state near the phase boundaries.

\begin{figure}[htbp]
\centering
\includegraphics[width=0.48\textwidth]{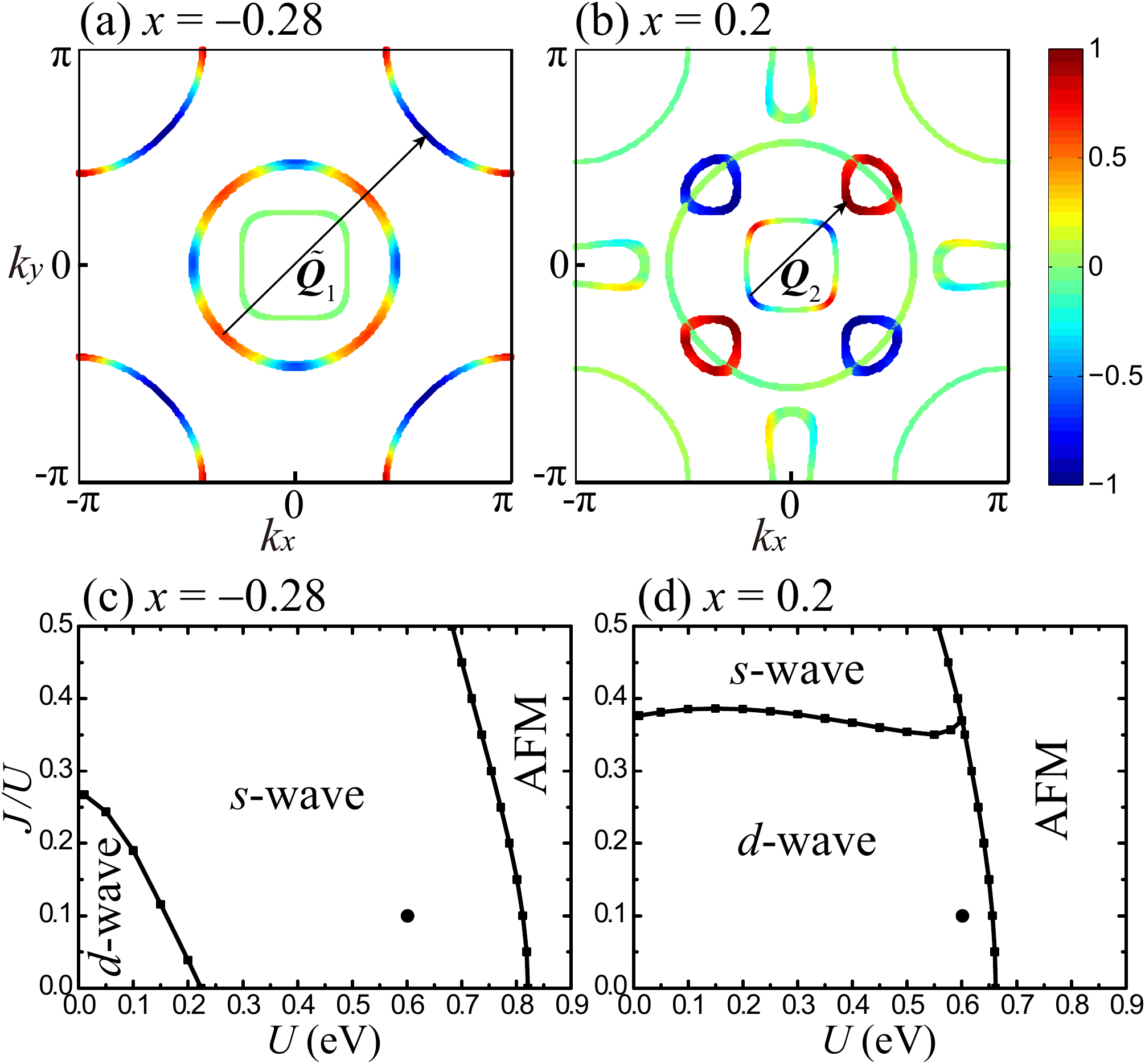}
\caption{(a) and (b) The gap functions of the leading pairing states with $U=0.6$ eV and $J/U=0.1$ under different doping levels $x$. $\tilde{\bm{Q}}_1(\approx \bm{Q}_1)$ and $\bm{Q}_2$ are the nesting vectors connecting the Fermi pockets. (c) and (d) The phase diagrams in $U$-$J/U$ plane under corresponding doping levels. The isolated black round dots indicate $U=0.6$ eV and $J/U=0.1$ adopted above.}\label{gap}
\end{figure}

To visualize the pairing symmetry of the leading $s$- and $d$-wave SC states, we depict their normalized gap functions at $x=-0.28$ and $0.2$ in Figs. \ref{gap}(a) and \ref{gap}(b), respectively. Obviously, the gap function of the $s$-wave pairing state has positive and negative changes at a single Fermi pocket, in addition to the symmetry about both $k_x$ and $k_y$ axes. Thus, more precisely, the SC state has the nodal $s_{\pm}$-wave pairing symmetry. On the other hand, the gap function of the $d$-wave pairing state is antisymmetric about both $k_x$ and $k_y$ axes. Thus, more precisely, the SC state has the $d_{xy}$-wave pairing symmetry. The further calculations at other doping levels can confirm that the leading $s$- and $d$-wave pairings in Fig. \ref{UcSC}(b) have the $s_{\pm}$- and $d_{xy}$-wave gap functions similar to those in Figs. \ref{gap}(a) and \ref{gap}(b), respectively.

Comparing Fig. \ref{gap}(a) [\ref{gap}(b)] with Fig. \ref{FSchi}(a) [\ref{FSchi}(c)], we notice that the $s(d)$-wave pairing develops mainly on the Fermi pockets with $d_{x^2-y^2}(d_{xz/yz})$-orbital character. If we depict the nesting vector $\tilde{\bm{Q}}_1$ ($\bm{Q}_2$) connecting the Fermi pockets with $d_{x^2-y^2}(d_{xz/yz})$-orbital character in Fig. \ref{gap}(a) [\ref{gap}(b)], we find that the gap function of the $s$($d$)-wave pairing satisfies $\text{sign}[\Delta_{\bm{k}}]=-\text{sign}[\Delta_{\bm{k}+\tilde{\bm{Q}}_1(\bm{Q}_2)}]$.
In fact, the gap function of the $d$-wave pairing satisfies this opposite sign condition at any two points on the $d_{xz/yz}$-orbital pockets connected by the nesting vectors including but not limited to $\bm{Q}_2$. This observation indicates that both $s$- and $d$-wave pairings are driven by the AFM spin fluctuations enhanced by the FS nestings.

Although the nesting mechanism applies to both $s$- and $d$-wave pairings, the nested Fermi pockets for these two pairings are completely decoupled in kinematics, and have very different doping dependence. Specifically, as shown in Figs. \ref{FSchi}(a)-\ref{FSchi}(c), the $d_{x^2-y^2}$-orbital pockets for the $s$-wave pairing change hardly as doping varies, and thus can always support the nesting. On the other hand, most $d_{xy}$-orbital pockets for the $d$-wave pairing change obviously as the doping varies, and have enough sizes to support the nesting only at relative high electron doping levels. This observation explains the crossover from the $s$-wave to the $d$-wave pairing as the doping level crosses the AFM range in the middle of Fig. \ref{UcSC}(b).

To test the robustness of the $s$- and $d$-wave SC states against the change of the interaction parameters, we focus on the doping levels $x=-0.28$ and $0.2$ to solve the gap equation (\ref{gapeq}) with different $U$ and $J/U$. For the hole doping case, as shown in Fig. \ref{gap}(c), the $s$-wave SC state dominates over the $d$-wave one in the large region of the SC part of the phase diagram, except in the region with small $U$ and $J/U$ where the $d$-wave pair becomes the dominant one. For the electron doping case, as shown in Fig. \ref{gap}(d), the $d$-wave SC state dominates over the $s$-wave one in the large region of the SC part of the phase diagram, except in the region with large $J/U$ where the $s$-wave pair becomes the dominant one. Since the $s$- and $d$-wave SC states are stable against the changes of the interaction parameters as well as the doping levels, we can expect that the results we presented here are suitable for the other Ni-based high-$T_c$ superconductors which share the similar single layer Ni$_2$M$_2$O (M=S, Se, Te).

\section{Conclusion}
To conclude, we have studied the susceptibility and pairing symmetry of newly predicted high-$T_c$ superconductor La$_2$Ni$_2$Se$_2$O$_3$, starting from the effective four-orbital TB model of its single [Ni$_2$Se$_2$O]$^{2-}$ layer. For the realistic interaction parameters, our calculations on the Hubbard-Hund model of the system reveal that the AFM state in parental compound is induced by the perfect FS nesting. Upon hole and electron dopings, we find that the compound hosts the $s_{\pm}$- and $d_{xy}$-wave pairing states, respectively. The $s$- and $d$-wave pairing states presented here can be distinguished by the phase-sensitive dc SQUID experiment, which has been used to determine the pairing symmetries in cuprates \cite{SQUID1} and Sr$_2$RuO$_4$ \cite{SQUID2}. The most interesting feature of the ground states of La$_2$Ni$_2$Se$_2$O$_3$ is the doping-type dependence of the pairing symmetry. This feature makes the system a platform to study the competition between the $s$- and $d$-wave SC states, and to reveal deeper similarities and differences between cuprates and iron-based superconductors.

\section*{Acknowledgements}
This work is supported by Beijing Natural Science Foundation (Grant No. 1174019) and the NSFC (Grant No. 11604013).


\begin{thebibliography}{*}

\bibitem{Hu1}
Hu J, Le C, Wu X. Predicting unconventional high-temperature superconductors in trigonal bipyramidal coordinations. Phys Rev X 2015;5:041012.


\bibitem{Hu2}
Hu J. Identifying the genes of unconventional high temperature superconductors. Sci Bull 2016;61:561.


\bibitem{Hu3}
Hu J, Le C. A possible new family of unconventional high temperature superconductors. Sci Bull 2017;62:212.


\bibitem{Hu4}
Le C, Qin S, Hu J. Electronic physics and possible superconductivity in layered orthorhombic cobalt oxychalcogenides. Sci Bull 2017;62:563.


\bibitem{Hu5}
Le C, Zeng J, Gu Y, Cao G-H, Hu J. A possible family of Ni-based high temperature superconductors. 2018;63:957.


\bibitem{RPA1}
Takimoto T, Hotta T, Ueda K. Strong-coupling theory of superconductivity in a degenerate Hubbard model. Phys Rev B 2004;69:104504.


\bibitem{RPA2}
Yada K, Kontani H. Origin of the weak pseudo-gap behaviors in Na$_{0.35}$CoO$_2$: absence of small hole pockets. J Phys Soc Jpn 2005;74:2161.


\bibitem{RPA3}
Kubo K. Pairing symmetry in a two-orbital Hubbard model on a square lattice. Phys Rev B 2007;75:224509.


\bibitem{RPA4}
Kuroki K, Onari S, Arita R, Usui H, Tanaka Y, Kontani H, Aoki H. Unconventional pairing originating from the disconnected Fermi surfaces of superconducting LaFeAsO$_{1-x}$F$_x$. Phys Rev Lett 2008;101:087004.


\bibitem{Scalapino1}
Graser S, Maier TA, Hirschfeld PJ, Scalapino DJ. Near-degeneracy of several pairing channels in multiorbital models for the Fe pnictides. New J Phys 2009;11:025016.


\bibitem{Scalapino2}
Maier TA, Graser S, Hirschfeld PJ, Scalapino DJ, $d$-wave pairing from spin fluctuations in the K$_x$Fe$_{2-y}$Se$_2$ superconductors. Phys Rev B 2011;83:100515(R).


\bibitem{Liu2013}
Liu F, Liu C-C, Wu K, Yang F, Yao Y. $d+id'$ chiral superconductivity in bilayer silicene. Phys Rev Lett 2013;111:066804.


\bibitem{Wu2014}
Wu X, Yuan J, Liang Y, Fan H, Hu J. $g$-wave pairing in BiS$_2$ superconductors. Europhys Lett 2014;108:27006.


\bibitem{Ma2014}
Ma T, Yang F, Yao H, Lin H. Possible triplet $p+ip$ superconductivity in graphene at low filling. Phys Rev B 2014;90:245114.


\bibitem{Zhang2015}
Zhang L-D, Yang F, Yao Y, Possible electric-field-induced superconducting states in doped silicene. Sci Rep 2015;5:8203.


\bibitem{Liu2018}
Liu C-C, Zhang L-D, Chen W-Q, Yang F. Chiral spin density wave and $d+id$ superconductivity in the the magic-angle-twisted bilayer graphene. Phys Rev Lett 2018;121:217001.


\bibitem{SQUID1}
Van Harlingen DJ. Phase-sensitive tests of the symmetry of the pairing state in
the high-temperature superconductors---Evidence for $d_{x^2-y^2}$ symmetry. Rev Mod
Phys 1995;67:515.


\bibitem{SQUID2}
Asano Y, Tanaka Y, Sigrist M, Kashiwaya S. Josephson interferometer in a
ring topology as a proof of the symmetry of Sr$_2$RuO$_4$. Phys Rev B 2005;71:214501.

\end{thebibliography}
\end{document}